\documentclass[A4paper,two column]{article}
\usepackage{graphicx}
\usepackage{amsmath}
\usepackage{amsfonts}

\begin{document}

\title{Precise detection of multipartite entanglement in four-qubit Greenberger--Horne--Zeilinger diagonal states}
\author{Xiao-yu Chen$^1$ \thanks{Email:xychen@zjgsu.edu.cn}, Li-zhen Jiang$^1$, Zhu-an Xu$^2$  \\
{\small {$^1$College of Information and Electronic Engineering, Zhejiang Gongshang University, Hangzhou, 310018, China}}\\
{\small {$^2$Department of Physics, Zhejiang University, Hangzhou, Zhejiang
310027, China }}}
\date{}
\maketitle

\begin{abstract}
We propose a method of constructing the separability criteria for multipartite quantum states on the basis of entanglement witnesses. The entanglement witnesses are obtained by finding the maximal expectation values of Hermitian operators and then optimizing over all possible Hermitian operators. We derive a set of tripartite separability criteria for the four-qubit Greenberger--Horne--Zeilinger (GHZ) diagonal states. The derived criterion set contains four criteria that are necessary and sufficient for the tripartite separability of the
highly symmetric four-qubit GHZ diagonal states; the criteria completely account for
the numerically obtained boundaries of the tripartite separable state set. One of the criteria is just the tripartite separability criterion of the four-qubit generalized Werner states.

PACS number(s): 03.65.Ud; 03.67.Mn;

\end{abstract}

\section{Introduction}

Multipartite entanglement is the main quantum resource in quantum computation, quantum simulation, and multipartite quantum communication. However, determining whether a given quantum state is multipartite entangled is a theoretically and experimentally challenging task. Many criteria have been developed to characterize and detect multipartite entanglement; see Ref.\cite{GuhneToth} for an overview. A solution to the entanglement detection problem, known as entanglement witnessing, relies on the geometry of the set of all separable quantum states\cite{Horodecki09},\cite{Horodecki96}. The entanglement witness (EW) method can easily be extended to multipartite cases\cite{Horodecki01}. Recent developments of the entanglement detection criteria are an EW for continuous variable systems\cite{Gerke}, an ultrafine EW\cite{Shahandeh1}, the semiquantum nonlocal game (SQNLG)\cite{Buscemi}, the EW game\cite{Zhou}, the relaxed nonlocality detection method\cite{Baccari}, the statistical speed\cite{Pezze}, and the separability eigenvalue equation (SEE)\cite{Sperling}. Most of them are based on EWs. In principle, there exist the extremal EW\cite{Sperling} and extremal SQNLG such that the entanglement criteria are necessary and sufficient. Practically, finding a solution to the SEE or finding an extremal SQNLG is still very difficult if not impossible.
Multipartite entanglement of a quantum state has rich structures\cite{Huber}. It has many levels of entanglement, which are usually characterized by the entanglement depth\cite{Sorensen}, i.e., the extent to which the quantum
state is many-body entangled. Genuine multipartite entanglement involving all parties of the system has the largest depth, and bipartite entanglement has the smallest depth.
All the other types of entanglement, with depths between these extremes, then can be called partial entanglement. The entanglement depth structure further complicates the detection and characterization of multipartite entanglement. EWs can be used to distinguish different classes of multipartite entanglement.
There is subtle difference between multipartite entanglement and multipartite separability. We will use multipartite separability to characterize quantum states in the following.

   The goal of this paper is to find a way to construct precise separability criteria for multipartite states. The proper starting point for this aim is to investigate the states that are diagonal in the Greenberger--Horne--Zeilinger (GHZ) basis\cite{Kay}.
GHZ diagonal states arise frequently as special multipartite quantum states in quantum information processing. They are tractable in many theoretical problems such as quantum channel capacity\cite{Chen2011}.
Most of the multipartite entangled states prepared in experiments are GHZ states. There have been recent experiments on four-qubit GHZ states; long-lived four-qubit GHZ states have been realized\cite{Kaufmann}, and a test of the irreducible four-qubit GHZ paradox has been produced\cite{Su}.
When imperfections in the preparation and decay are considered, the prepared states are usually GHZ diagonal states. The precise relationship between the positive partial transpose (PPT) criterion and full separability of GHZ diagonal states has been studied\cite{Kay}, and a simple condition has been given for the equivalence of the PPT criterion and full separability.
When the condition is not fulfilled, the boundary between full separability and entanglement cannot be determined using the PPT criterion.
Then a complicated EW should be devised to detect the boundary. For three-qubit GHZ diagonal states, an EW has been found\cite{GuhnePLA}\cite{Chen2015}\cite{Chen2017};
hence, the necessary and sufficient criterion of full separability is known. Research on the multipartite entanglement of GHZ diagonal states has concentrated on the criteria of biseparability and full separability\cite{GuhneSeevinck}. To be concrete while still considering the rich structure of multipartite entanglement, we will consider the problem of tripartite separability, which is neither the problem of biseparability nor that of full separability, for four-qubit GHZ diagonal states.

\section{Preliminary}
Suppose there is a composed Hilbert space $\mathcal{H}=\mathcal{H}_{1}\otimes\cdot\cdot\cdot\otimes\mathcal{H}_{n}$.
Consider a partition $\mathcal{I}=\{\mathcal{I}_{1},..., \mathcal{I}_{k}\}$ of the index set $\mathcal{J} = \{1, . . . , n\}$. A quantum state $\sigma_{\mathcal{I}}$ is called separable for the given partition $\mathcal{I}$ if it can be written as a classical mixture of product states:
\begin{equation}\label{1a}
\sigma_{\mathcal{I}}=\sum_{i}q_{i}|\psi^{(i)}_{\mathcal{I}_{1}}\rangle\langle\psi^{(i)}_{\mathcal{I}_{1}}|\otimes\cdot\cdot\cdot\otimes|\psi^{(i)}_{\mathcal{I}_{k}}\rangle\langle\psi^{(i)}_{\mathcal{I}_{k}}|,
\end{equation}
where $q_{i}$ is a classical probability distribution, and $|\psi^{(i)}_{\mathcal{I}_{j}}\rangle$ is a pure state of subset $\mathcal{I}_{j}$. A state $\sigma$ is called $k$-separable if it can be written as
\begin{equation}\label{1b}
 \sigma=\sum_{\mathcal{I}:|\mathcal{I}|=k}q'_{\mathcal{I}}\sigma_{\mathcal{I}},
\end{equation}
where $|\mathcal{I}|$ is the number of elements in the set $\mathcal{I}$, and $q'_{\mathcal{I}}$ is a classical probability distribution. The summation is over all possible $k$-partite partitions. If a quantum state cannot be written in the form of Eq.(\ref{1b}), it is referred to as $k$-inseparable. A $2$-inseparable (not biseparable) state is also called genuinely entangled.

A four-qubit GHZ diagonal state takes the form
\begin{equation}\label{1c}
  \rho=\sum_{j=1}^{16}p_{j}|GHZ_{j}\rangle\langle GHZ_{j}|,
\end{equation}
where $p_{j}$ is a probability distribution. The GHZ state basis consists of 16 vectors, $|GHZ_{j}\rangle=\frac{1}{\sqrt{2}}(|0x_{2}x_{3}x_{4}\rangle \pm|1\overline{x}_{2}\overline{x}_{3}\overline{x}_{4}\rangle)$, where $x_{i}, \overline{x}_{i}\in\{0, 1\}$, and $x_{i} \neq \overline{x}_{i}$. In
binary notation, $j-1 = 0x_{2}x_{3}x_{4}$ for the ``+'' states, and $j-1 = 1\overline{x}_{2}\overline{x}_{3}\overline{x}_{4}$ for the ``-'' states.

The highly symmetric four-qubit GHZ diagonal state investigated in this paper, which takes the form
   \begin{eqnarray}\label{1d}
     \rho=p_{1}|GHZ_{1}\rangle\langle GHZ_{1}|+p_{16}|GHZ_{16}\rangle\langle GHZ_{16}|\nonumber\\
     +p_{2}\sum_{j=2}^{8}|GHZ_{j}\rangle\langle GHZ_{j}|+p_{15}\sum_{j=9}^{15}|GHZ_{j}\rangle\langle GHZ_{j}|,
   \end{eqnarray}
is a special GHZ diagonal state with $p_{i}\geq0$ and normalization
\begin{equation}\label{1e}
     p_{1}+p_{16}+7(p_{2}+p_{15})=1.
\end{equation}

A generalized Werner state (a GHZ state mixed with white noise\cite{Pittenger}),
\begin{equation}\label{1f}
     \rho_{W}=p|GHZ\rangle\langle GHZ|+\frac{1-p}{16}\mathbf{I},
\end{equation}
is a special highly symmetric GHZ diagonal state, where $|GHZ\rangle=|GHZ_{1}\rangle$, and $\mathbf{I}$ is the $16\times 16$ identity matrix.

An EW is a Hermite operator $\hat{W}$ such that $Tr\rho_{s}\hat{W}\geq 0$ for all separable states $\rho_{s}$
($k$-separable or separable in some given sense) and $Tr\rho\hat{W}<0$ for at least one entangled state $\rho$ (with a certain associated entanglement depth). We may assume $\hat{W}=\Lambda\mathbb{I}-\hat{M}$,
where $\mathbb{I}$ is the identity operator, and $\Lambda=\max_{\rho_s}Tr\rho_s\hat{M}$ such that
$\hat{W}$ is an optimal EW (the equality in $Tr\rho_{s}\hat{W}\geq 0$ can be reached). We may express the multi-qubit state and the EW using their characteristic functions.
Thus, the operator $\hat{M}$ is characterized by real parameters $M_{i}$ $(i=1,...,4^{n}-1)$ in detecting the entanglement of an $n$-qubit state.
Here the number of parameters $M_{i}$ is equal to the number of free real parameters for describing the density matrix.
One of the widely used numerical methods of finding a proper EW employs semidefinite programming.
The procedure of analytically finding a precise EW is divided into two steps. The first step is to find $\Lambda$ for the given $M_{i}$.
Notice that any operator $\hat{M}$ corresponds to a valid EW if $\Lambda$ is obtained. Hence, the first step gives a valid necessary criterion of separability.
The second step is to adjust the parameters $M_{i}$ such that the EW detects all the entanglement of a given depth. The parameters $M_{i}$ should match the state under consideration,
so the second step gives the sufficient criterion of separability.

The two steps for finding the entanglement criteria are just the two types of optimization. The first step is the maximization to obtain $\Lambda$ (and thus the optimal EW)
for a given set of parameters $M_{i}$. The second step is optimization with respect to $M_{i}$ such that the criterion is tight.
Randomly choosing the parameters $M_{i}$ will lead to very inefficient optimization in the two steps. Thus, the problem is how to choose $M_{i}$ properly. Given the GHZ diagonal states of (\ref{1c}), we may assume $M_{i}$ ($i=1,...,15$) as free parameters. It is very difficult to treat $15$ parameters analytically. Therefore, we seek some symmetries to reduce the number of parameters $M_{i}$. Then we derive the EWs by two optimization steps. It follows that some necessary criteria of entanglement for the GHZ diagonal states of (\ref{1c}) can be obtained. To check whether the criteria are also sufficient, we explicitly decompose the states of (\ref{1d}), which are special states of (\ref{1c}), into tripartite separable states. We prove that the set of criteria is not sufficient for general states of (\ref{1c}), and some other criteria should be found to realize sufficiency. On the other hand, we find that one of the criteria we obtained suffices as the necessary and sufficient criterion for the states of (\ref{1f}), which are special states of (\ref{1d}).

We describe the framework of the criteria as follows. Let $\mathcal{S}_{1}$,$\mathcal{S}_{2}$,...,$\mathcal{S}_{N}$ be a hierarchy of state sets such that
$\mathcal{S}_{1}\subset\mathcal{S}_{2}\subset\cdot\cdot\cdot\subset\mathcal{S}_{N}$. Let $\mathcal{C}_{i}$ be the set of necessary and sufficient entanglement criteria for state set $\mathcal{S}_{i}$. Notice that $\mathcal{S}_{1}\subset\mathcal{S}_{2}$, so the criterion set $\mathcal{C}_{1}$
detects at least some of the states in $\mathcal{S}_{2}$ necessarily and sufficiently. We have $\mathcal{C}_{1}\subseteq\mathcal{C}_{2}$, and eventually $\mathcal{C}_{1}\subseteq\mathcal{C}_{2}\subseteq\cdot\cdot\cdot\subseteq\mathcal{C}_{N}$. Thus, for a larger state set, we may add additional criteria to the criterion set.
Each criterion set $\mathcal{C}_{i}$ can be converted to an EW set $\mathcal{W}_{i}$, where $\mathcal{W}_{i}$ can detect the entanglement of $\mathcal{S}_{i}$ necessarily and sufficiently.
Hence, we have $\mathcal{W}_{1}\subseteq\mathcal{W}_{2}\subseteq\cdot\cdot\cdot\subseteq\mathcal{W}_{N}$.

To detect the multipartite entanglement of multi-qubit systems (the state set $\mathcal{S}_{N}$), we should consider a hierarchy of state sets.
The sets can be the graph-diagonal state set $\mathcal{S}_{4}$, GHZ diagonal-state set $\mathcal{S}_{3}$, highly symmetric GHZ diagonal-state set $\mathcal{S}_{2}$, or generalized Werner state set $\mathcal{S}_{1}$. We will show that $\mathcal{C}_{1}\subset\mathcal{C}_{2}\subset\mathcal{C}_{3}$.
The generalized Werner state set has been considered for the four-qubit case,
and all the parameters $M_{i}$ have been determined\cite{Chen2017}. The necessary and sufficient criterion of tripartite separability has been given\cite{Chen2017}. We will recognize the criterion obtained\cite{Chen2017} from the criterion set $\mathcal{C}_{2}$ and denote it as the criterion set $\mathcal{C}_{1}$. We will build the criterion set $\mathcal{C}_{2}$, which is necessary and sufficient for the state set $\mathcal{S}_{2}$ and is a fairly good necessary criterion set for larger state sets $\mathcal{S}_{j}$ with $j\geq3$. Further, we find that the two EW sets $\mathcal{W}_{1}$ and $\mathcal{W}_{2}$ share some common parameters $M_{i}$.

\section{Optimal entanglement witness}\label{section2}
Let $\hat{M}$ be a Hermitian operator that is a linear combination of the tensor products of the Pauli operators appearing in the four-qubit GHZ diagonal states, namely,
\begin{eqnarray}\label{1}\nonumber
 \hat{M}&=&M_{1}IIZZ+M_{2}IZIZ+M_{3}IZZI+M_{4}ZIIZ\\
&&+M_{5}ZIZI+M_{6}ZZII\nonumber+M_{7}ZZZZ \\
&&+M_{8}XXXX+M_{9}XXYY\nonumber+M_{10}XYXY\\
&&+M_{11}XYYX+M_{12}YXXY+M_{13}YXYX\nonumber \\
&&+M_{14}YYXX+M_{15}YYYY,
\end{eqnarray}
where $X,Y,Z$ are Pauli matrices, $I$ is the $2\times2$ identity matrix, and  $M_{i}$ are the parameters mentioned above. For partition $1|2|34$ (more formally, partition $\mathcal{I}=\{\mathcal{I}_{1},\mathcal{I}_{2},\mathcal{I}_{3}\}$ with $\mathcal{I}_{1}=1,\mathcal{I}_{2}=2$, and $\mathcal{I}_{3}=\{3,4\}$; the four qubits are divided into three parties, and the third party has two qubits), the mean of the operator $\hat{M}$ on
the pure product state $|\psi\rangle=|\psi_{1}\rangle|\psi_{2}\rangle|\psi_{34}\rangle$ is $\langle\psi|\hat{M}|\psi\rangle$,
where $|\psi_{1}\rangle,|\psi_{2}\rangle$, and $|\psi_{34}\rangle$ are the pure states of the first,
second, and last two qubits, respectively. We may alternatively express the mean as $\langle\psi_{34}|\mathcal{M}|\psi_{34}\rangle$,
where $\mathcal{M}=\langle\psi_{1}|\langle\psi_{2}|\hat{M}|\psi_{1}\rangle|\psi_{2}\rangle$ is a $4\times4$ matrix.
For a given $|\psi_{1}\rangle,|\psi_{2}\rangle$, we can maximize the mean $\langle\psi|\hat{M}|\psi\rangle$ as the largest
eigenvalue of the matrix $\mathcal{M}$. For GHZ diagonal states, the structure of the matrix $\mathcal{M}$ is of the ``X'' type.
The matrix contains diagonal and antidiagonal entries, and all the other entries are zeros.
Thus, the candidates for the largest eigenvalue of $\mathcal{M}$ are easily obtained as
\begin{eqnarray}\label{2}
  \lambda_{1}=\frac{1}{2}(\mathcal{M}_{11}+\mathcal{M}_{44}+\sqrt{(\mathcal{M}_{11}-\mathcal{M}_{44})^2+4|\mathcal{M}_{14}|^2}),\\
  \lambda_{2}=\frac{1}{2}(\mathcal{M}_{22}+\mathcal{M}_{33}+\sqrt{(\mathcal{M}_{22}-\mathcal{M}_{33})^2+4|\mathcal{M}_{23}|^2}),
\end{eqnarray}
where $\mathcal{M}_{mn}$ are the entries of the matrix $\mathcal{M}$. Let the Bloch vectors of pure states $|\psi_{i}\rangle$ ($i$ = 1,2)
be $(\sin\theta_{i}\cos\varphi_{i},\sin\theta_{i}\sin\varphi_{i},\cos\theta_{i})$. We have
\begin{eqnarray}\label{3}\nonumber
\mathcal{M}_{11}=K_{0}+K_{2}\cos\theta_{2}+ K_{4}\cos\theta_{1}+K_{6}\cos\theta_{1}\cos\theta_{2},\nonumber\\
\mathcal{M}_{44}=K_{0}-K_{2}\cos\theta_{2}- K_{4}\cos\theta_{1}+K_{6}\cos\theta_{1}\cos\theta_{2},\nonumber\\
\mathcal{M}_{22}=K_{1}+K_{3}\cos\theta_{2}+ K_{5}\cos\theta_{1}+K_{7}\cos\theta_{1}\cos\theta_{2},\nonumber\\
\mathcal{M}_{33}=K_{1}-K_{3}\cos\theta_{2}- K_{5}\cos\theta_{1}+K_{7}\cos\theta_{1}\cos\theta_{2},\nonumber
\end{eqnarray}
where $(K_{0},K_{2},K_{4},K_{6})=(M_{1},M_{2}+M_{3},M_{4}+M_{5},M_{6}+M_{7})$, $(K_{1},K_{3},K_{5},K_{7})=(-M_{1},M_{2}-M_{3},M_{4}-M_{5},M_{6}-M_{7})$,
and
 $\mathcal{M}_{14}=\sin\theta_{1}\sin\theta_{2}g_{1}(\varphi_{1},\varphi_{2}),$ $\mathcal{M}_{23}=\sin\theta_{1}\sin\theta_{2}g_{2}(\varphi_{1},\varphi_{2}),$ with
\begin{eqnarray}\label{4}\nonumber
g_{1(2)}(\varphi_{1},\varphi_{2})=K_{8(9)}\cos\varphi_{1}\cos\varphi_{2}-iK_{10(11)}\cos\varphi_{1}\sin\varphi_{2}\\
-iK_{12(13)}\sin\varphi_{1}\cos\varphi_{2}+K_{14(15)}\sin\varphi_{1}\sin\varphi_{2},\nonumber
\end{eqnarray}
where $(K_{8},K_{10},K_{12},K_{14})=(M_{8}-M_{9},M_{10}+M_{11},M_{12}+M_{13},M_{14}-M_{15})$, and
$(K_{9},K_{11},K_{13},K_{15})=(M_{8}+M_{9},M_{10}-M_{11},M_{12}-M_{13},M_{14}+M_{15})$.

   The maximization of the mean of the operator $\hat{M}$ over the partition $1|2|34$ is reduced to
   maximization with respect to the four angles $\theta_{i}$,$\varphi_{i}$ ($i$ = 1,2). We can see that
   the maximization on $\varphi_{i}$ is independent of the maximization on $\theta_{i}$.
   Let $\widetilde{g}_{j}=\max_{\varphi_{1},\varphi_{2}}|g_{j}(\varphi_{1},\varphi_{2})|$ ($j$ = 1,2);
   then we have the following result for $\widetilde{g}_{j}$ (the proof can be found in Appendix A).

   \begin{equation}\label{5}
     \widetilde{g}_{j}=\left\{\begin{array}{c}
               \sqrt{\frac{(\xi\beta+\gamma\delta)(\xi\gamma+\beta\delta)(\xi\delta+\beta\gamma)}{\xi\beta\gamma\delta}}, \text{ if }   \xi\beta\gamma\delta>0  \text{ and }  \widetilde{q}\geq0;\nonumber\\
               \max_{i=j+7,j+9,j+11,j+13} |K_{i}|,   \text{  otherwise},
             \end{array}
             \right.
   \end{equation}
   where  $(\xi,\beta,\gamma,\delta)=\frac{1}{4}(K_{8(9)},K_{10(11)},K_{12(13)},K_{14(15)})\Gamma$, with $\Gamma$
   being a $4\times 4$ matrix in which all of the diagonal entries are $-1$ and the off-diagonal entries are $+1$.
   Here $\widetilde{q}=q_{0}q_{1}q_{2}q_{3}$, where $(q_{0}q_{1}q_{2}q_{3})=(\xi\beta\gamma,\xi\beta\delta,\xi\gamma\delta,\beta\gamma\delta)\Gamma.$

   Let $\widetilde{\lambda}_{j}=\max_{\varphi_{1},\varphi_{2}}\lambda_{j}$; then
   \begin{eqnarray}\label{6}\nonumber
     \widetilde{\lambda}_{j}&=&K_{j-1}+K_{j+5}\cos\theta_{1}\cos\theta_{2}\\
&+&\sqrt{(K_{j+1}\cos\theta_{1}+K_{j+3}\cos\theta_{2})^2+(\widetilde{g}_{j}\sin\theta_{1}\sin\theta_{2})^2}\nonumber.
   \end{eqnarray}
   For general parameters $K_{i} $ and $\widetilde{g}_{j}$, it is not obvious how $\theta_{i}$ ($i$ = 1,2) can be removed
   from the expression of the eigenvalue by maximization. A practical way is to guess the proper parameters and then
   check whether the EW is the correct one and the criterion is sufficient for separability.
   To simplify the problem, we consider the symmetry of the parameter $M_{i}$ under different partitions.
   Different partitions can be obtained from existing ones by interchanging qubits: $1|2|34\Leftrightarrow 1|3|24$
   if qubits $ 2\Leftrightarrow 3$. As a result, $M_{i}$, and thus $K_{i} $, will also be interchanged.
   We assume
   \begin{eqnarray}
    M_{1}=M_{2}=M_{3}=M_{4}=M_{5}=M_{6},\label{6x}\\
    M_{9}=M_{10}=M_{11}=M_{12}=M_{13}=M_{14}.\label{6y}
   \end{eqnarray}
   The assumption may limit the entanglement detection power of the optimal EW derived from the operator $\hat{M}$.
   However, it greatly simplifies the analysis. The maximal eigenvalue is already the maximal mean of $\hat{M}$ for all partitions
   (and thus the tripartite separability) by symmetry. The symmetric assumption of the parameters $M_{i}$
   leads to $K_{3}=K_{5}=K_{11}=K_{13}=0$; then $\widetilde{\lambda}_{2}$ is
   \begin{eqnarray}\label{6a}\nonumber
     \widetilde{\lambda}_{2}&=&K_{1}+K_{7}\cos\theta_{1}\cos\theta_{2}+\widetilde{g}_{2}|\sin\theta_{1}\sin\theta_{2}|\\
     &\leq& K_{1}+\max(|K_{7}|,\widetilde{g}_{2}),
   \end{eqnarray}
   where
   \begin{equation}\label{6e}
     \widetilde{g}_{2}=\max(|K_{9}|,|K_{15}|).
   \end{equation}
For the particular tripartite separable problem of four-qubit GHZ diagonal states, we further assume

 \begin{equation}\label{6z}
 M_{1}=0,
 \end{equation}
 and thus $K_{i}=0 (i=0,...,5)$, and $K_{6}=-K_{7}=M_{7}$. Let $\Lambda_{i}=\max_{\theta_{1},\theta_{2}}\widetilde{\lambda}_{i}$; then
 \begin{eqnarray}
     \Lambda_{1}&=&\max(|M_{7}|,\widetilde{g}_{1})\label{6b},\\
     \Lambda_{2}&=&\max(|M_{7}|,\widetilde{g}_{2})\label{6c}.
   \end{eqnarray}
The maximal mean of $\hat{M}$ over all possible tripartite separable states is
   \begin{equation}\label{6d}
     \Lambda=\max(\Lambda_{1},\Lambda_{2})=\max(|M_{7}|,\widetilde{g}_{1},\widetilde{g}_{2}).
   \end{equation}

The present parameters $M_{i}$ are compatible with and closely related to the EW for the triseparability of generalized Werner states.
The generalized Werner state EW has $M_{1}=M_{2}=M_{3}=M_{4}=M_{5}=M_{6}=0$, $M_{7}=2$, $M_{8}=M_{15}=1$, and $M_{9}=M_{10}=M_{11}=M_{12}=M_{13}=M_{14}=-1$\cite{Chen2017}. Assumptions (\ref{6x}), (\ref{6y}), and (\ref{6z}) on the parameters $M_{i}$
reveal the relationship between the two sets of EWs; namely, $\mathcal{W}_{1}$ and $\mathcal{W}_{2}$ share some common $M_{i}$.

\section{Matched entanglement witness}
   The optimal EW can be constructed as $\hat{W}=\Lambda \mathbb{I}-\hat{M}$. If $\hat{W}$
   can detect the entanglement of a state $\rho$, then Tr$\rho\hat{W}<0$; namely, $Tr\rho\hat{M}>\Lambda$.
   A four-qubit GHZ diagonal state can be written as
   \begin{eqnarray}\label{7}\nonumber
 \rho&=&\frac{1}{16}(IIII+R_{1}IIZZ+R_{2}IZIZ+R_{3}IZZI\\
&&+R_{4}ZIIZ+R_{5}ZIZI+R_{6}ZZII+R_{7}ZZZZ\nonumber\\
&&+R_{8}XXXX+R_{9}XXYY+R_{10}XYXY\nonumber\\
&&+R_{11}XYYX+R_{12}YXXY+R_{13}YXYX\nonumber \\
&&+R_{14}YYXX+R_{15}YYYY).
\end{eqnarray}
   Then $Tr\rho\hat{M}=\sum_{i=1}^{15}M_{i}R_{i}=\sum_{i=1}^{15}K_{i}T_{i}$, with
   $T_{2i},T_{2i+1}=\frac{1}{2}(R_{2i}\pm R_{2i+1})$ for $i$ = 0,1,2,3,5,6 (we assume $R_{0}=0$ here), and
    $T_{2i},T_{2i+1}=\frac{1}{2}(R_{2i}\mp R_{2i+1})$ for $i$ = 4,7. Let
   \begin{equation}\label{8}
     \mathcal{L}=\frac{\Lambda}{\sum_{i=1}^{15}K_{i}T_{i}}.
   \end{equation}
   Using the convention $\sum_{i=1}^{15}K_{i}T_{i}>0$, we say that the entanglement of $\rho$ is detected if $\mathcal{L}<1$. For all possible optimal EWs,
   we want to find an EW with the smallest $\mathcal{L}$. We will call it the matched EW with respect to the given state $\rho$.
   \begin{equation}\label{9}
     \mathcal{L}_{min}=\min_{\hat{M}}\mathcal{L}.
   \end{equation}

   To minimize $\mathcal{L}$ with respect to $K_{i}$ ($i$ = 1,...,15), we first consider
     $\widetilde{R}_{1}=\max\sum_{i=8,10,12,14}K_{i}T_{i}/\widetilde{g}_{1}$. We have\cite{Chen2017}
   \begin{equation}\label{10}
     \widetilde{R}_{1}=\left\{\begin{array}{l}
                           \sqrt{\frac{(T_{8}T_{10}+T_{12}T_{14})(T_{8}T_{12}+T_{10}T_{14})(T_{8}T_{14}+T_{10}T_{12})}{T_{8}T_{10}T_{12}T_{14}}}\\ \text{ \ \qquad    \qquad  for  } T_{8}T_{10}T_{12}T_{14}>0 \text{  and   } Q\geq 0 \\
                           \max_{i=8,10,12,14}{|t_{i}|}, \text{ otherwise},
                           \end{array}
                           \right.
   \end{equation}
where $Q=Q_{0}Q_{1}Q_{2}Q_{3}$, with $(Q_{0},Q_{1},Q_{2},Q_{3})=T_{8}T_{10}T_{12}T_{14}(\frac{1}{T_{8}},\frac{1}{T_{10}},\frac{1}{T_{12}},\frac{1}{T_{14}})\Gamma$;
$(t_{8},t_{10},t_{12},t_{14})=(T_{8},T_{10},T_{12},T_{14})\Gamma$.

Let $\widetilde{R}_{2}=\max\sum_{i=9,15}K_{i}T_{i}/\widetilde{g}_{2}=|T_{9}|+|T_{15}|.$ When $\widetilde{R}_{1}\geq\widetilde{R}_{2}$, the minimization of $\mathcal{L}$ is simplified to $\mathcal{L}_{min}=\min_{\widetilde{g}_{2}\leq \widetilde{g}_{1}}\mathcal{\widetilde{L}}$, with
\begin{eqnarray}\nonumber
   \mathcal{\widetilde{L}}&=&\min_{M_{7},\widetilde{g}_{1}}\frac{\Lambda}{M_{7}R_{7}+\widetilde{g_{1}}\widetilde{R}_{1}+\widetilde{g_{2}}\widetilde{R}_{2}}\\
   &=&\frac{1}{|R_{7}|+\widetilde{R}_{1}+\widetilde{R}_{2}\widetilde{g}_{2}/\Lambda}\label{11}.
\end{eqnarray}
We may leave the minimization with respect to $\widetilde{g}_{2}$ pending. Alternatively, when $\widetilde{R}_{2}\geq\widetilde{R}_{1}$, we have $\mathcal{L}_{min}=\min_{\widetilde{g}_{1}\leq \widetilde{g}_{2}}\mathcal{\widetilde{L'}}$, with
\begin{equation}
   \mathcal{\widetilde{L'}}=\frac{1}{|R_{7}|+\widetilde{R}_{2}+\widetilde{R}_{1}\widetilde{g}_{1}/\Lambda}\label{11a}.
\end{equation}
When $\widetilde{g}_{2}$=$\widetilde{g}_{1}$, we would obtain $\mathcal{L}_{min}$. However,
care must be taken when we write the expression $\mathcal{L}_{min}$.
The reason is that $\widetilde{g}_{1}$ and $\widetilde{g}_{2}$ may not be independent;
they are correlated owing to our assumptions on the parameters $M_{i}$. The details can be found in Appendix B.
When $M_{9}=0$, we have $\widetilde{g}_{2}$=$\widetilde{g}_{1}$. For $M_{9}=0$,
$\widetilde{R}_{1}$ is reduced to $|T_{8}|+|T_{14}|$ instead of expression (\ref{10}).
 The correct result should be
\begin{equation}\label{12}
  \mathcal{L}_{min}=\frac{1}{|R_{7}|+|R_{8}|+|R_{15}|}.
\end{equation}

\section{Separability criterion}
The separability criteria are $\widetilde{\mathcal{L}}\geq 1$ and $\mathcal{L}_{min}\geq 1 $.
The separability criteria for the partition $1|2|34$ derived from (\ref{11}) is
\begin{equation}
  |R_{7}|+\widetilde{R}_{1}+x\widetilde{R}_{2}\leq1,\label{14}\\
\end{equation}
where $x\in [0,1)$. When $x=0$, we have the separability criterion
\begin{equation}\label{16}
  |R_{7}|+\widetilde{R}_{1} \leq1.
\end{equation}
An interesting case appears for criterion (\ref{16}) when $\widetilde{{R}}_{1}$ is equal to its second line in (\ref{10}).
With the entries of $\rho$, we have $t_{8}=-8\rho_{1,16},t_{10}=8\rho_{4,13}, t_{12}=8\rho_{5,12}$, and $t_{14}=8\rho_{8,9}$.
Hence, when conditions $T_{8}T_{10}T_{12}T_{14}>0$ and $Q>0$ are not fulfilled, we have the triseparability criterion.
\begin{eqnarray}\label{16a}
\text{Criterion I:\ \qquad  \qquad \qquad  \qquad} \nonumber\\
\max(|\rho_{1,16}|,|\rho_{4,13}|,|\rho_{5,12}|,|\rho_{8,9}|)\leq\frac{1}{2}\min(\rho_{1,1}+\rho_{4,4}\nonumber\\
+\rho_{6,6}+\rho_{7,7},\text{\quad}\rho_{2,2}+\rho_{3,3}+\rho_{5,5}+\rho_{8,8}).
\end{eqnarray}
Criterion (\ref{16a}) is a necessary criterion for tripartite separability. The criterion derived from (\ref{12}) is

\begin{equation}
\text{Criterion II:   }|R_{7}|+|R_{8}|+|R_{15}|\leq1. \label{15}
\end{equation}
Criteria (\ref{16a}) and (\ref{15}) give rise to the linear boundaries (straight lines in Figs.1 and 2)
of the tripartite separable state set. We anticipate that the nonlinear boundaries (curves in the figures)
are attributed to the first line of formula (\ref{10}). However, this is not the case, at least for the example
 in the next section. There is another way of minimizing $\mathcal{L}$ in (\ref{9}). We may write
 $\sum_{i=1}^{15}M_{i}R_{i}=M_{7}R_{7}+M_{8}R_{8}+M_{9}\sum_{i=9}^{14} R_{i}+M_{15}R_{15}$ owing to our assumptions
  on $M_{i}$. We may rewrite it as $M_{7}R_{7}+\sum_{i=8,10,12,14}K_{i}T'_{i}$, where $T'_{8}=R_{8}$,
  $T'_{10}=T'_{12}= \frac{1}{4}(\sum_{i=8}^{15}R_{i})$, and $T'_{14} =-R_{15}$. We then have the tripartite separability criterion
\begin{equation}\label{17a}
  |R_{7}|+\widetilde{R'}_{1}\leq 1,
\end{equation}
where $\widetilde{R'}_{1}$ is defined just as $\widetilde{R}_{1}$ in (\ref{9}), with $T_{i}$ being
replaced by $T'_{i}$. We simplify the separability criterion (\ref{17a}) to
  \begin{eqnarray}\label{17b}
  \text{Criterion III:\ \qquad  \qquad \qquad  \qquad} \nonumber\\
    1-|R_{7}|\geq |R_{8}-R_{15}|\sqrt{1-\frac{(\sum_{i=8}^{15}R_{i})^2}{16R_{8}R_{15}}}
\end{eqnarray}
 if $R_{8}R_{15}<0$ and $|8R_{8}R_{15}|\geq|(\sum_{i=8}^{15}R_{i}) (R_{8}+R_{15})|$.

 We can see that the above case corresponds to $\widetilde{g}_{2}<\widetilde{g}_{1}$. There is also the
  case $\widetilde{g}_{2}=\widetilde{g}_{1}$. It can be obtained by setting

 \begin{equation}\label{17c}
   \widetilde{g}_{1}=|M_{7}|=\max(|M_{8}+M_{9}|,|M_{9}+M_{15}|).
 \end{equation}
 The number of free parameters is reduced from three, $(M_{8},M_{9},M_{15})$, to two owing to
 condition (\ref{17c}). Notice that we may fix one of them (see $M_{9}$) without affecting
 $\mathcal{L}$. Hence, there is only one free parameter (see $M_{15}$) left for minimizing $\mathcal{L}$.
 We then have the separability criterion
 \begin{eqnarray}\label{17d}
 \text{Criterion IV:\ \qquad  \qquad  \qquad} \nonumber\\
   |R_{7}|+\widetilde{R''}\leq 1,
 \end{eqnarray}
 where $\widetilde{R''}=\min_{M_{15}}\frac{\sum_{i=8}^{15}M_{i}R_{i}}{|M_{7}|}$
subject to (\ref{17c}),
 and $ M_{i}=M_{9}=-1 (i=10,...,14)$.

We have shown that there are four criteria for the triseparability of four-qubit GHZ diagonal states.
We will show that these criteria are necessary and sufficient for the triseparability of highly symmetric four-qubit GHZ diagonal states.
We may denote the criterion set as $\mathcal{C}_{2}=$ \{ criterion I, criterion II, criterion III, criterion IV\}. Let us denote $\mathcal{C}_{1}=$ \{ Criterion I\}.
Because criterion I is the necessary and sufficient criterion of triseparability for generalized Werner states (see Appendix C), we have $\mathcal{C}_{1}\subset\mathcal{C}_{2}$.

\section{Application to highly symmetric GHZ diagonal states}
\begin{figure}[tpb]
\includegraphics[ trim=0.000000in 0.000000in -0.138042in 0.000000in,
height=2.5in, width=3.5in]{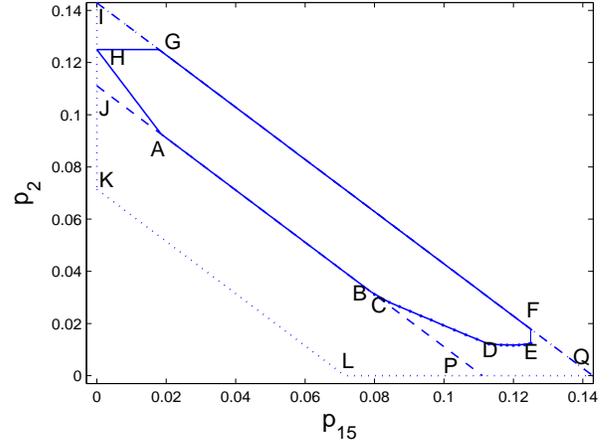}
\caption{(1) Solid lines: Numerically calculated triseparable state set of highly symmetric four-qubit GHZ diagonal states with $p_{16}=0.$ Straight line FG is the physical--unphysical boundary.
 Criterion I accounts for straight line AB. Criterion II accounts for straight lines EF, GH, and AH. Criterion III accounts for curve BC. Criterion IV accounts for curve CDE. (2) Dashed lines: Necessary tripartite separable set determined by criterion I and physical boundaries, shown as IJPQI. (3) Dotted lines: Biseparable state set IKLQI. }
\end{figure}

As defined in Eq.(\ref{1d}), a four-qubit highly symmetric GHZ diagonal state $\rho$ is a mixture of GHZ basis states with the probabilities \{$ p_{i},i=1,...,16$\} and $p_{2}=p_{3}=...=p_{8}$; $p_{9}=p_{10}=...=p_{15}$.
The state is symmetric under interchange of any pair of qubits.
 Hence, we obtain four positive parameters, $p_{1},p_{2},p_{15}$, and $p_{16}$, with the normalization
\begin{equation}\label{18}
  p_{1}+p_{16}+7(p_{2}+p_{15})=1.
\end{equation}
The nonzero entries of $\rho$ are $\rho_{1,1}=\rho_{16,16}=\frac{1}{2}(p_{1}+p_{16})$;
$\rho_{1,16}=\rho_{16,1}=\frac{1}{2}(p_{1}-p_{16})$; $\rho_{2,2}=\rho_{3,3}=...=\rho_{15,15}=\frac{1}{2}(p_{2}+p_{15})$; and
$\rho_{2,15}=\rho_{3,14}=...=\rho_{15,2}=\frac{1}{2}(p_{2}-p_{15})$.

We numerically calculated the boundaries of the tripartite separable state sets for the above states
for $p_{16}=0$ and $p_{16}=0.3$. We chose $p_{2}$ and $p_{15}$ as free parameters,
and $p_{1}$ was determined by the normalization in (\ref{18}). The boundaries are shown in Figs.1 and 2.
The numerical calculation has rounds of three steps: (i) choose $M_{i}$ randomly, (ii) calculate $\Lambda$, and (iii) record the minimal $\mathcal{L}$.
   Let $p_{15}=\frac{v}{q},p_{2}=\frac{1-v}{q}$, and $v\in[0,1]$. The normalization in (\ref{18}) gives the
   upper bound of $1/q$. Because $p_{1}=1-p_{16}-7(p_{2}+p_{15})$, we have
   \begin{equation}\label{19}
     p_{2}+p_{15}\leq \frac{1}{7}(1-p_{16}).
   \end{equation}
The equality in (\ref{19}) gives the straight-line boundary FG in Fig.1, with $q=\frac{7}{1-p_{16}}=7$.

For convenience, we list the relevant $R_{j}$ below:
\begin{eqnarray}
  R_{7} &=&R_{1}=1-8(p_{2}+p_{15}), \label{19a}\\
  R_{8} &=& 1-2p_{16}-14p_{15}, \label{19b}\\
  R_{15}&=&-R_{9}=1-2p_{16}-8p_{2}-6p_{15}. \label{19c}
\end{eqnarray}

\begin{figure}[tpb]
\includegraphics[ trim=0.000000in 0.000000in -0.138042in 0.000000in,
height=2.5in, width=3.5in]{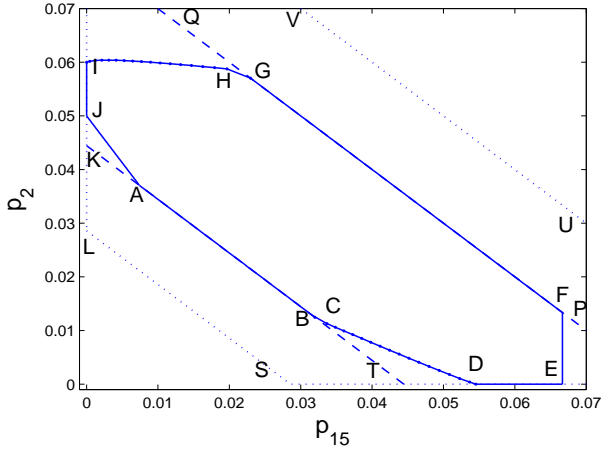}
\caption{(1) Solid lines: Numerically calculated triseparable state set of highly symmetric four-qubit GHZ diagonal states with $p_{16}=0.3$.
Criterion I accounts for straight lines AB and FG. Criterion II accounts for straight lines DE, EF, IJ, and AJ. Criterion III accounts for curves BC and GH. Criterion IV accounts for curves CD and HI. (2) Dashed lines: Necessary triseparable state set using criterion I and physical boundaries (the boundaries are line section KT, line $p_{15}=0$, line $p_{2}=0$, and line PQ). (3)Dotted lines: Biseparable state set (the boundaries are line section LS, line $p_{15}=0$, line $p_{2}=0$, and line UV).}
\end{figure}

\subsection{Necessary criteria}
    Except for the physical--unphysical boundary determined by (\ref{19}), all the other boundaries in Figs.1 and 2 are related to the necessary criteria of tripartite separability.
    For our state $\rho$, criterion I is
    \begin{equation}\label{20}
      |p_{1}-p_{16}|\leq 2(p_{2}+p_{15}).\\
    \end{equation}
    When $p_{1}>p_{16}$, the criterion gives a lower bound of $1/q$ corresponding to the straight-line boundary AB
    in Figs.1 and 2, with $q=9$ and $q=22.5$, respectively. When $p_{1}<p_{16}$, the criterion gives an upper
    bound of $1/q$ corresponding to the straight-line boundary FG in Fig.2, with $q=12.5$.

    Criterion II can be written as $R_{8}+R_{15}+R_{7}\leq 1$, $-R_{8}+R_{15}-R_{7}\leq 1$,
     $-R_{8}+R_{15}+R_{7}\leq 1$, $R_{8}-R_{15}-R_{7}\leq 1$, $R_{8}-R_{15}+R_{7}\leq 1$, and $-R_{8}-R_{15}+R_{7}\leq 1$ for different cases. They are
    \begin{eqnarray}
      4p_{2}+7p_{15}&\geq& \frac{1}{2}-p_{16}, \label{21} \\
      p_{15}&\leq& \frac{1}{8}, \label{22} \\
      p_{2}&\geq& 0, \label{22a}\\
      p_{2}&\leq& \frac{1}{8}, \label{23} \\
      p_{15}&\geq& 0, \label{23a}\\
      3p_{15}&\leq& \frac{1}{2}-p_{16}, \label{24}
    \end{eqnarray}
    respectively. Inequality (\ref{21}) accounts for the straight-line boundaries AH in Fig.1 and AJ in Fig.2.
    Inequality (\ref{22}) accounts for the straight-line boundary EF in Fig.1. Inequality (\ref{23}) accounts
    for the straight-line boundary GH in Fig.1. Inequality (\ref{24}) accounts for the straight-line boundary EF in Fig.2.
    The straight-line boundaries DE and IJ in Fig.2 are described by inequalities (\ref{22a}) and (\ref{23a}), respectively.
    The conditions in (\ref{22a}) and (\ref{23a}) are also the physical--unphysical boundaries.

    What is left are the boundary curves BCDE in Fig.1 and BCD and GHI in Fig.2. The sections BC and GH
    can be explained by criterion III. The sections CDE in Fig.1 and CD and HI in Fig.2 are
    explained by criterion IV.

    We denoted the intersections of the criteria as points A, B, ... in Figs.1 and 2.
    Let $\alpha=\frac{q}{(1-2p_{16})}$. Then
    \begin{equation}\label{24a0}
    (p_{15},p_{2})=(\frac{v(1-2p_{16})}{\alpha},\frac{(1-v)(1-2p_{16})}{\alpha}).
    \end{equation}
     We have $\alpha=9$ for the straight line AB in either Fig.1 or 2, $v\in [v_{A},v_{B}]$, with

\begin{equation}\label{24a}
      v_{A}=\frac{1}{6}; \text{    } v_{B}=\frac{1}{16}(5+\sqrt{41})\simeq 0.7127.
\end{equation}
We obtain $v_{A}=\frac{1}{6}$ from (\ref{20}) and (\ref{21}). The value of $v_{B}$ is the
result of the equality in criterion III. Because $R_{7}=1-\frac{8(1-2p_{16})}{\alpha}$, we have $R_{7}>0$ for
points B and G in Fig.2 and also for point B in Fig.1. The equality in criterion III can be reduced to a power equation of $v$:
\begin{equation}\label{24b}
  \sum_{i=0}^{4}a_{i}v^{i}=0,
\end{equation}
 where $a_{4}=256, a_{3}=96\alpha-1184,$ $a_{2}=1364-144\alpha, a_{1}=60\alpha-520$, and $a_{0}=(\alpha-10)^2$.
 Equation (\ref{24b}) determines the curves BC and GH in Fig.2 and the curve BC in Fig.1. For $\alpha=9$,
 the power equation is $(16v^2-10v-1)^2=0$, which gives the solution $v_{B}$ in (\ref{24a}).
 Furthermore, for $\alpha=5$, the power equation is $(16v^2-22v-5)^2=0$. The solution
 $v=v_{G}=\frac{1}{16}(11-\sqrt{41})\simeq 0.2873$ determines the location of point G in Fig.2.
 We have $\alpha=5$ for the straight line FG in Fig.2 with $v\in [v_{F},v_{G}]$. Further, $v_{F}=\frac{5}{6}$
 comes from (\ref{20}) and (\ref{24}). We can see that $v_{G}=1-v_{B}$ and  $v_{F}=1-v_{A}$ in Fig.2.
 The power equation in (\ref{24b}) is invariant under the transformation $\alpha-7\Rightarrow 7-\alpha, v\Rightarrow 1-v$.
 Thus, we need to analyze only the curve BC instead of both BC and GH in Fig.2.

 The curves CD and HI in Fig.2 and CDE in Fig.1 correspond to separability criterion IV.
 To simplify the notation, let $s=1-\frac{M_{8}+M_{15}}{2M_{9}}$ and $t=\frac{M_{15}-M_{8}}{2M_{9}}$.
 The condition in (\ref{17c}) leads to
 \begin{equation}\label{24c}
   |t|=(1-\frac{4}{s^2})(s-2)-\frac{4}{s^2}\sqrt{(1-s)(4-s^2)}.
 \end{equation}
Let $K=-\frac{R_{8}}{R_{15}}$; then $\widetilde{R''}=R_{15}\tau(K)$, with
\begin{equation}\label{24d}
\tau(K)=\max_{s\in[0,1]}\frac{s(1-K)+|t|(1+K)+K+5}{|s-|t|-2|}.
\end{equation}

Criteria III and IV intersect at point C (or H). In Fig.3, we show that the
boundary in $(\alpha,v)$ coordinates changes from criterion III to criterion IV when $\alpha$ decreases.
The exact coordinates of C should be determined. Criterion IV becomes $R_{15}\tau(K)\leq 1-R_{7}$. When $R_{7}>0$,
we have $1-R_{7}=\frac{8}{q}=\frac{8}{\alpha}(1-2p_{16})$, and $R_{15}=(1-\frac{8-2v}{\alpha})(1-2p_{16})$.
The equality in criterion IV is simplified to $(\alpha-8+2v)\tau(K)=8$. $K$ is defined as $K=-\frac{K_{8}}{K_{15}}=\frac{14v-\alpha}{\alpha-8+2v}$.
Hence, the boundary curve CD in both Figs.1 and 2 is determined by the following parameter equation:
\begin{equation}\label{24e}
  v=\frac{1}{2}[1+\frac{K+1}{\tau(K)}], \text{ \qquad} \alpha=7+\frac{7-K}{\tau(K)}. \\
\end{equation}
The curve CD (dot-dashed) is tangent to the curve BC (solid) in Fig.3 at point C. Combining Eq.(\ref{24e}) with
power equation (\ref{24b}) gives the coordinates of point C: $v_{C}=0.7492394,\alpha_{C}=8.900032.$ The curves HI in Fig.2 and DE in Fig.1 can be analyzed similarly.

The numerical curves CD and HI in Fig.2 and CDE in Fig.1 are obtained by a random search of the EW operators. The parameters for points C and H obtained by the random search are $v_{C}=0.7492,
\alpha_{C}=8.90, v_{H}=0.2508, \alpha_{H}=5.50$. In Fig.2, D is the end point of criterion IV
with $\alpha_{D}=7.3333$. In Fig.1, E is the end point of criterion IV with $\alpha_{E}=7.273, v_{E}=0.9091$. The numerical curves fit
separability criterion IV well.

\begin{figure}[tpb]
\includegraphics[ trim=0.000000in 0.000000in -0.138042in 0.000000in,
height=2.5in, width=3.5in]{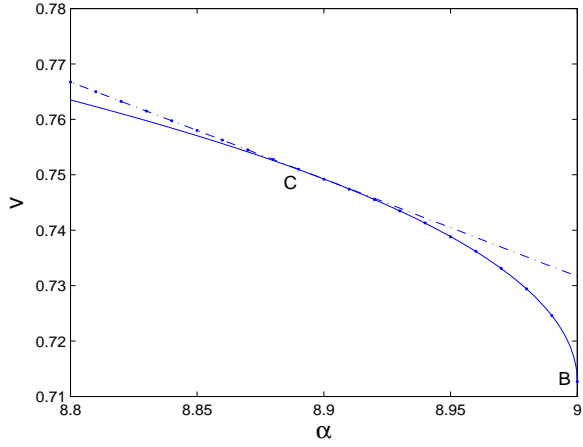}
\caption{Intersection of criteria III and IV at point C.
The solid curve represents criterion III, which is not valid for $\alpha<\alpha_{C}$.
 The dash-dotted curve represents criterion IV, which is not valid for $\alpha>\alpha_{C}$.}
\end{figure}
    \subsection{Sufficient criteria}\label{AA}

    The sufficient condition of separability relies on the ability to decompose the state into a probability
    mixture of product states. It is usually rather technically complex to write the decomposition. For our known
    operator $\mathcal{M}$, we will find the eigenvectors corresponding to its largest eigenvalue $\Lambda$ and
    use these eigenvectors to construct the explicit decomposition of a state $\rho$ at the boundary.
    \subsubsection{Sufficiency of criterion I}
    The choice of $M_{1}=0,M_{7}=1$ and $M_{8}=M_{15}=-M_{9}=\frac{1}{2}$ leads to criterion I. The matrix $\mathcal{M}$ is
    \begin{equation}\label{25}\nonumber
      \mathcal{M}=\left(
                          \begin{array}{cccc}
                            c_{1}c_{2}& 0 & 0 & s_{1}s_{2}e^{i(\varphi_{1}+\varphi_{2})} \\
                            0 & -c_{1}c_{2} & 0 & 0 \\
                            0 & 0 &-c_{1}c_{2}& 0 \\
                            s_{1}s_{2}e^{-i(\varphi_{1}+\varphi_{2})} & 0 & 0 & c_{1}c_{2} \\
                          \end{array}
                        \right),
    \end{equation}
    where $c_{i}=\cos\theta_{i}$, and $s_{i}=\sin\theta_{i}.$ The largest eigenvalue of $\mathcal{M}$ is $\Lambda=1$. The corresponding eigenvector is
    \begin{equation}\label{26}
      |\psi_{34}\rangle=\frac{1}{\sqrt{2}}(|00\rangle+ e^{-i(\varphi_{1}+\varphi_{2})}|11\rangle).
    \end{equation}
    The tripartite product state can be given by $|\psi(\varphi_{1},\varphi_{2})\rangle=|\psi_{1}\rangle|\psi_{2}\rangle|\psi_{34}\rangle$,
    where $|\psi_{k}\rangle=\frac{1}{\sqrt{2}}(|0\rangle+e^{i\varphi_{k}}|1\rangle),k=1,2$, and we have set $\theta_1=\theta_2=0$ for simplicity. Let
    $\varrho(\varphi_{1},\varphi_{2})$ =$|\psi(\varphi_{1},\varphi_{2})\rangle\langle\psi(\varphi_{1},\varphi_{2})|$.
    Further, let $\varrho_{1}(\varphi_{1},\varphi_{2})$ =$\frac{1}{8}\sum_{k_{0},k_{1},k_{2}=0}^{1} \varrho((-1)^{k_{0}}(\varphi_{1}+k_{1}\pi),(-1)^{k_{0}}(\varphi+k_{2}\pi))$ be a mixture of $\varrho$ for different angles $\varphi_{k}$. The state $\varrho_{1}$ is tripartite separable with the partition $1|2|34$.

     Averaging over all six partitions, we then have a tripartite separable state
    $\overline{\varrho}_{1}=\frac{1}{6}\sum_{j=1}^{6}\varrho_{1}(j)$ (where $j$ denotes different partitions). The explicit $\overline{\varrho}_{1}$ is shown in Appendix D, and we set $\varphi_{\pm}=\varphi_{1}\pm\varphi_{2}$ there.
The highly symmetric GHZ diagonal state on the tripartite separable boundary (straight lines AB in Figs.1 and 2 and FG in Fig.2) can be expressed as
    \begin{equation}\label{27a}
      \rho=(1-\kappa)\rho'+\kappa\overline{\varrho}_{1},
    \end{equation}
where $\rho'$ is a fully separable state, and $\kappa=\frac{4|\alpha-7|(1-2p_{16})}{\alpha}$ (see Appendix D).
    The antidiagonal part of state $\rho$ [compared with Eq.(\ref{7})] is characterized by
    \begin{eqnarray}
      R_{8}&=&\frac{\kappa}{2}(\cos^2\varphi_{+}+\cos\varphi_{+}\cos\varphi_{-}),\label{28a}\\
      R_{15}&=&\frac{\kappa}{2}(\cos^2\varphi_{+}-\cos\varphi_{+}\cos\varphi_{-}),\label{28b}\\
      R_{9}&=&-\frac{\kappa}{6}(1+\sin^2\varphi_{+}).\label{28c}
    \end{eqnarray}
 For the four-qubit highly symmetric GHZ diagonal states, we have $R_{9}+R_{15}=0$. This leads to
 \begin{equation}\label{29}
   \cos\varphi_{+}=\frac{1}{8}(3\cos\varphi_{-}\pm\sqrt{9\cos^2\varphi_{-}+32}).
 \end{equation}
 Consequently, $\cos\varphi_{+} \in ([-1,-\frac{\sqrt{41}-3}{8}]\cup [\frac{\sqrt{41}-3}{8},1])$.
 We thus have $\frac{R_{8}}{R_{15}}=\frac{7\cos^2\varphi_{+}-2}{2-\cos^2\varphi_{+}}$.
 On the other hand, we have $\frac{R_{8}}{R_{15}}=\frac{\alpha-14v}{\alpha-8+2v}$ from (\ref{19b}), (\ref{19c}), and (\ref{24a0}). Thus,
 \begin{equation}\label{30}
   v=\frac{1}{6}[\alpha-4+2(7-\alpha)\cos^2\varphi_{+}].
 \end{equation}
Hence, $v\in [\frac{1}{6}, \frac{1}{16}(5+\sqrt{41})]$ when $\alpha=9$ (straight line AB in Fig.1 or 2).
 Namely, $v_{A}=\frac{1}{6}$, and $v_{B}=\frac{1}{16}(5+\sqrt{41})$.
Further, $v\in [\frac{1}{16}(11-\sqrt{41}),\frac{5}{6}]$ when $\alpha=5$ (straight line FG in Fig.2).
Namely, $v_{F}=\frac{5}{6}$, and $v_{G}=\frac{1}{16}(11-\sqrt{41})$.
 These results are exactly the same as those obtained by the necessary criteria.
  Thus, the straight lines AB in Figs.1 and 2 and FG in Fig.2 are exact
  boundaries of the tripartite separable state set.

\subsubsection{Sufficiency of criterion II}
 The choice of $M_{7}=\pm1,M_{8}=\pm1,M_{9}=0$, and $M_{15}=\pm1$ leads to criterion II.
 First, we consider the case $M_{8}=M_{15}=1$. The matrix $\mathcal{M}$ is
\begin{equation}\label{31}\nonumber
      \mathcal{M}=\left(
                          \begin{array}{cccc}
                            c_{1}c_{2}& 0 & 0 & s_{1}s_{2}b_{+} \\
                            0 & -c_{1}c_{2} & s_{1}s_{2}b_{-} & 0 \\
                            0 & s_{1}s_{2}b_{-} &-c_{1}c_{2}& 0 \\
                            s_{1}s_{2}b_{+} & 0 & 0 & c_{1}c_{2} \\
                          \end{array}
                        \right),
    \end{equation}
    where $b_{\pm}=\cos\varphi_{\pm}$, and $M_{7}=1$ is assumed. The four eigenvalues of $\mathcal{M}$ reach their maximal value of 1 under different conditions.
    These conditions are (i) $\theta_{1}-\theta_{2}=0, \varphi_{+}=0$, (ii) $\theta_{1}-\theta_{2}=0, \varphi_{+}=\pi$,
    (iii) $\theta_{1}-\theta_{2}=\pi, \varphi_{-}=0$, and (iv) $\theta_{1}-\theta_{2}=\pi, \varphi_{-}=\pi$ for the four eigenvalues, respectively.
    We consider case (i) first. The corresponding eigenvector is
    \begin{equation}\label{32}
      |\psi_{34}\rangle=\frac{1}{\sqrt{2}}(|00\rangle+|11\rangle).
    \end{equation}
    The product state is $|\psi(\theta,\varphi)\rangle=|\psi_{1}\rangle|\psi_{2}\rangle|\psi_{34}\rangle$,
    where $|\psi_{1}\rangle=\cos\frac{\theta}{2}|0\rangle+\sin\frac{\theta}{2}e^{i\varphi}|1\rangle$, and $|\psi_{2}\rangle=\cos\frac{\theta}{2}|0\rangle+\sin\frac{\theta}{2}e^{-i\varphi}|1\rangle.$
    We have defined $\theta=\theta_{1}$ and $\varphi=\varphi_{1}$.
    Let $\varrho(\theta,\varphi)=|\psi(\theta,\varphi)\rangle\langle\psi(\theta,\varphi)|$. We define
    \begin{equation}\label{33}
      \varrho_{2}(\theta,\varphi)=\frac{1}{4}[\varrho(\theta,\varphi)+\varrho(-\theta,\varphi)+\varrho(\theta+\pi,\varphi)+\varrho(-\theta+\pi,\varphi)].
    \end{equation}
    The state $\varrho_{2}(\theta,\varphi)$ comes from the first eigenvalue of $\mathcal{M}$.
    Similarly, we have the other three states, $\varrho_{i}(\theta,\varphi)$ ($i$ = 3,4,5), derived
    from the other three eigenvalues of $\mathcal{M}$.
    Let $\varrho_{23}(\theta,\varphi)=\frac{1}{2}\sum_{i=2}^{3}\varrho_{i}(\theta,\varphi)$ and
     $\varrho_{45}(\theta,\varphi)=\frac{1}{2}\sum_{i=4}^{5}\varrho_{i}(\theta,\varphi)$.
      The states are derived from the partition $1|2|34$. Averaging over all six partitions, we have $\overline{\varrho}_{23(45)}(\theta,\varphi)=
       \frac{1}{6}\sum_{j=1}^{6}\varrho_{23(45)}(\theta,\varphi,j)$ (where $j$ denotes different partitions); see Appendix D.

    Let the constructed tripartite separable GHZ diagonal state be
    \begin{equation}\label{34a}
    \rho=p\overline{\varrho}_{23}(\theta,\varphi)+(1-p)\overline{\varrho}_{45}(\theta,\varphi'),
    \end{equation}
    which is a probability mixture of the two states. Comparing the state $\rho$ with Eq.(\ref{7}), we have
    \begin{eqnarray}
      R_{1} &=&\frac{1}{6}(2p-1)(1+\cos^2\theta),  \text{  } R_{7}=\cos^2\theta,\label{35} \\
      R_{8} &=& \sin^2\theta(p\cos^2\varphi+(1-p)\cos^2\varphi'), \\
      R_{9} &=& -\frac{1}{6}\sin^2\theta(2p-1),\\
      R_{15} &=& \sin^2\theta(p\sin^2\varphi+(1-p)\sin^2\varphi').
    \end{eqnarray}
    Using $R_{9}+R_{15}=0$ (this requires $p\geq \frac{1}{2}$), we
     arrive at $\frac{R_{8}}{R_{15}}=\frac{7-2p}{2p-1}$. Notice that
     $\frac{R_{8}}{R_{15}}=\frac{\alpha-14v}{\alpha-8+2v}$; then
    \begin{equation}\label{36}
      v=\frac{1}{6}[\alpha-4+\frac{2(7-\alpha)}{p}].
    \end{equation}
    The equality in (\ref{21}) can be written as $\alpha=8+6v$ (this is the
    equation of the straight lines AJ in Fig.2 and AH in Fig.1). Substituting $\alpha=8+6v$ into
    Eq.(\ref{36}), we obtain $v=\frac{1}{6}(2p-1)$.
    Thus, we have $v\in [0,\frac{1}{6}]$ for the straight lines AJ and AH.
    Once again, we obtain $v_{A}=\frac{1}{6}$. From (\ref{19a}) and (\ref{35}), we have $\cos^2\theta =\frac{2p-1}{7-2p}\in [0,\frac{1}{5}]$. Notice that $R_{7}+R_{8}+R_{15}=1$; the triseparable state $\rho$ in (\ref{34a}) is on the boundary of triseparability.

We then choose $M_{8}=M_{15}=-1$ and $M_{7}=1$, and construct the GHZ diagonal state; Eq.(\ref{36}) remains true. The EW corresponds to the necessary condition in (\ref{24}).
The equality in condition (\ref{24}) can be written as $\alpha=6v$. Substituting it into
Eq.(\ref{36}), we have $v=\frac{1}{6}(7-2p)$.
Hence, we have $v\in [\frac{5}{6},1]$ for the straight line EF in Fig.2. Once again, we find
 that the coordinates of point F in Fig.2 are $v_{F}=\frac{5}{6}, \alpha_{F}=5$.

For $M_{8}=-1$ and $M_{15}=1$, the separability criteria are (\ref{22})
when $M_{7}=-1$ and (\ref{22a}) when $M_{7}=1$, which correspond to the straight lines EF in
Fig.1 and IJ in Fig.2. The constructed GHZ diagonal state is
 different from that for $M_{8}=M_{15}=\pm1$. We have
 \begin{eqnarray}\label{37}
      R_{8} &=&-\sin^2\theta(p\cos^2\varphi+(1-p)\cos^2\varphi'), \\
      R_{9} &=& \frac{1}{6}\sin^2\theta[p\cos(2\varphi)-(1-p)\cos(2\varphi')],\\
      R_{15} &=& \sin^2\theta(p\sin^2\varphi+(1-p)\sin^2\varphi').
 \end{eqnarray}
 Using $R_{9}+ R_{15}=0$, we can express $p$ as a function of $\sin^2\varphi$ and $\sin^2\varphi'$.
 We calculate the minimum of $K\equiv-\frac{R_{8}}{R_{15}}$ with respect to $\sin^2\varphi$ and $\sin^2\varphi'$.
 Then we find $K\geq 5$. On the other hand, $K=-\frac{\alpha-14v}{\alpha-8+2v}$. The equation for line EF in Fig.1
 is $p_{15}=\frac{1}{8}$; namely, $\alpha=8v(1-2p_{16})=8v$ (notice that $p_{16}=0$ in Fig.1). Thus, $v=\frac{4K}{5K-3}\leq \frac{10}{11}$.
  We find that the coordinates of E in Fig.1 are $v_{E}=\frac{10}{11}\simeq 0.9091, \alpha_{E}=\frac{80}{11}\simeq 7.2727 $,
  in perfect agreement with the numerical result. The equation for line IJ in Fig.2 is $p_{2}=0$; namely, $v=0$.
  We have $K=\frac{\alpha}{8-\alpha}\geq 5$; then $\alpha \in [\frac{20}{3},8]$ or $\alpha_{I}=\frac{20}{3}, \alpha_{J}=8$. Because $p_{2}=\frac{1-2p_{16}}{\alpha}$ for
  line IJ in Fig.2, $p_{2}=0.06$ for point I, and $p_{2}=0.05$ for point J. The sufficient condition is in complete agreement with the necessary condition.

  The choice of $M_{8}=1$ and $M_{15}=-1$ leads to separability criterion (\ref{23}) when $M_{7}=-1$ and separability criterion (\ref{23a}) when $M_{7}=1$.
  They correspond to lines GH in Fig.1 and DE in Fig.2, respectively. We can obtain the triseparable GHZ diagonal state
  with $K=-\frac{R_{8}}{R_{15}}\geq 5$ as in the previous situation. Thus, $-\frac{\alpha-14v}{\alpha-8+2v}\geq 5$. The equation for line GH in Fig.1 is
  $p_{2}=\frac{1}{8}$, or $\alpha=8(1-v)(1-2p_{16})$ (notice that $p_{16}=0$ in Fig.1). Therefore, $v\leq \frac{2}{13}$. This is consistent with $v_{G}=\frac{1}{8}$.
  Point G is in the triseparable state set. In the line $p_{2}=\frac{1}{8}$, the points with $v>\frac{1}{8}$ are nonphysical,
  although they do not conflict with $v\leq \frac{2}{13}$. The equation for the line DE in Fig.2 is $p_{15}=0$; namely, $v=1$. We have $-\frac{\alpha-14}{\alpha-6}\geq 5$;
  thus, $\alpha\in [6,\frac{22}{3}]$. The sufficient criterion coincides with the necessary criterion.

  We have omitted the analyses of $R_{1}$ and $R_{7}$ except for criterion (\ref{21}).
\subsubsection{Sufficiency of criteria III and IV}
For criteria III and IV, the matrix $\mathcal{M}$ takes the form
\begin{equation}\label{38}\nonumber
      \mathcal{M}=\left(
                          \begin{array}{cccc}
                            M_{7}c_{1}c_{2}& 0 & 0 & s_{1}s_{2}g_{1} \\
                            0 & -M_{7}c_{1}c_{2} & s_{1}s_{2}g_{2} & 0 \\
                            0 & s_{1}s_{2}g^{*}_{2} &-M_{7}c_{1}c_{2}& 0 \\
                            s_{1}s_{2}g^{*}_{1} & 0 & 0 & M_{7}c_{1}c_{2} \\
                          \end{array}
                        \right).
\end{equation}
We have shown that the maximum of $|g_{1}(\varphi_{1},\varphi_{2})|$ is $\widetilde{g}_{1}$. Suppose that the maximum is reached
at $\varphi_{1}=\phi_{1}$ and $\varphi_{2}=\phi_{2}$. Let $g_{1}(\phi_{1},\phi_{2})=\widetilde{g}_{1}e^{i\phi_{3}}.$ Then the largest
eigenvalue of $\mathcal{M}$ is $\widetilde{\lambda}_{1}=\widetilde{g}_{1}$ (we have chosen $M_{7}=\widetilde{g}_{1}$) with the eigenvector
\begin{equation}\label{39}
  |\psi_{34}\rangle=\frac{1}{\sqrt{2}}(|00\rangle+e^{-i\phi_{3}}|11\rangle).
\end{equation}
The properties for $\phi_{3}$ are as follows: $\phi_{1}\rightarrow \phi_{1}+\pi$ leads to $\phi_{3}\rightarrow \phi_{3}+\pi$, and
 $\phi_{1},\phi_{2}\rightarrow -\phi_{1},-\phi_{2}$ leads to $\phi_{3}\rightarrow -\phi_{3}$. The product state is $|\psi\rangle=|\psi_{1}\rangle|\psi_{2}\rangle|\psi_{34}\rangle,$
 where $|\psi_{j}\rangle=\cos\frac{\theta}{2}|0\rangle+\sin\frac{\theta}{2}e^{i\phi_{j}}|1\rangle$ ($j$ = 1,2).
 To construct the separable state according to criterion III, let $\varrho=|\psi\rangle\langle\psi|$,
 $\text{  } \varrho_{6}(\theta)=\frac{1}{4}[\varrho(\theta)+\varrho(-\theta)+\varrho(\theta+\pi)++\varrho(-\theta-\pi)]$, and $\text{  } \varrho_{7}(\phi_{1},
 \phi_{2},\phi_{3})=\frac{1}{4}[\varrho_{6}(\phi_{1},\phi_{2},\phi_{3})+\varrho_{6}(\phi_{1}+\pi,\phi_{2},\phi_{3}+\pi)
 +\varrho_{6}(-\phi_{1},-\phi_{2},-\phi_{3})+\varrho_{6}(-\phi_{1}+\pi,-\phi_{2},-\phi_{3}+\pi)]$; we have omitted $\theta$ or $\phi_{i}$ when doing so does not cause confusion.
 Then
 \begin{eqnarray}\label{40}\nonumber
   \varrho_{7}=\frac{1}{16}\{(II+\cos^2\theta ZZ)(II+ZZ) \\
   +\sin^2\theta[(r_{0}XX+r_{3}YY)(XX-YY)\nonumber \\
   +(r_{1}XY+r_{2}YX)(XY+YX)]\},
 \end{eqnarray}
 where $r_{0}=\cos\phi_{1}\cos\phi_{2}\cos\phi_{3}$,   $r_{1}=\cos\phi_{1}\sin\phi_{2}\sin\phi_{3},$
   $r_{2}=\sin\phi_{1}\cos\phi_{2}\sin\phi_{3},$ and  $r_{3}=\sin\phi_{1}\sin\phi_{2}\cos\phi_{3}$. Averaging over all six partitions, $\overline{\varrho}_{7}=\frac{1}{6}\sum_{j=1}^{6}\varrho_{7}(j)$ (where $j$ denotes different partitions),
   we have the triseparable state $\rho=\overline{\varrho}_{7}$ with
   \begin{eqnarray}\label{41}
     R_{7}=\cos^2\theta,\text{  } R_{8}=\sin^2\theta r_{0},\\
     R_{9}=\frac{1}{6}\sin^2\theta(-r_{0}+r_{3}+2r_{1}+2r_{2}),\\
     R_{15}=-\sin^2\theta r_{3}.
   \end{eqnarray}
 The variables $r_{i}$ are related as follows: $(r_{0}r_{1}+r_{2}r_{3})(r_{0}r_{2}+r_{2}r_{3})(r_{0}r_{3}+r_{1}r_{2})=r_{0}r_{1}r_{2}r_{3}$. If we set $r_{1}=r_{2}$,
  this reduces to
  \begin{equation}\label{42}
    (r_{0}+r_{3})^2(r_{0}r_{3}+r_{1}^2)=r_{0}r_{3}.
  \end{equation}
   Namely,
  \begin{equation}\label{43}
    \sin^2\theta=|R_{8}-R_{15}|\sqrt{1-\frac{(R_{8}+R_{15}+6R_{9})^2}{16R_{8}R_{15}}}\\
    =\widetilde{R'}_{1}.
  \end{equation}
  Notice that $R_{7}=\cos^2\theta$; thus, we arrive at $R_{7}+\widetilde{R'}_{1}=1$.
 Therefore, the states represented by curves BC in both Figs.1 and 2 and GH in Fig.2 are triseparable. Criterion III is necessary and sufficient.

  The eigenvalue $\widetilde{\lambda}_{2}$ may also contribute to the separable state with its eigenvectors
  $|\psi_{34}\rangle=|01\rangle$ and $|\psi_{34}\rangle=|10\rangle$ when $\widetilde{g}_{2}<\widetilde{g}_{1}$.
  These terms can modify only the diagonal components of the density matrix of the separable state. However, the eigenvector of $\widetilde{\lambda}_{2}$ does contribute to the
  antidiagonal components of the density matrix (in the computational basis) when $\widetilde{g}_{2}=\widetilde{g}_{1}$. This is the case for criterion IV. We have the eigenvector
  $|\psi_{34}\rangle=\frac{1}{\sqrt{2}}(|01\rangle+e^{-i\varphi_{3}}|10\rangle)$ (where $\varphi_{3}=0,\pi$) for eigenvalue $\widetilde{\lambda}_{2}$.
  The product state is $|\psi\rangle=|\psi_{1}\rangle|\psi_{2}\rangle|\psi_{34}\rangle$, where
  $|\psi_{1}\rangle=\cos\frac{\theta}{2}|0\rangle+\sin\frac{\theta}{2}|1\rangle$, and
  $|\psi_{2}\rangle=-\sin\frac{\theta}{2}|0\rangle+\cos\frac{\theta}{2}e^{i\varphi_{3}}|1\rangle$. The constructed triseparable state is
  \begin{equation}\label{44}
    \varrho_{8}=\frac{1}{16}[(II-\cos^2\theta ZZ)(II-ZZ)-\sin^2\theta XX(XX+YY)].
  \end{equation}
Averaging over all the partitions, we have the triseparable state $\overline{\varrho}_{8}$.
Mixing this state with the state $\overline{\varrho}_{7}$, we obtain the triseparable state $\rho=(1-\eta)\overline{\varrho}_{7}+\eta\overline{\varrho}_{8}$, where
 \begin{eqnarray}\label{45}
     R_{8}=\sin^2\theta [(1-\eta)r_{0}-\eta],\\
     R_{9}=\frac{1}{6}\sin^2\theta[(1-\eta)(-r_{0}+r_{3}+4r_{1})-\eta],\\
     R_{15}=-\sin^2\theta (1-\eta)r_{3},\\
     R_{7}=\cos^2\theta.
   \end{eqnarray}
   Let $R'_{8}=\frac{R_{8}}{1-|R_{7}|},R'_{15}=\frac{R_{8}}{1-|R_{7}|}$, and $R'_{9}=\frac{R_{8}+6R_{9}+R_{15}}{4(1-|R_{7}|)}$;
   then Eq.(\ref{42}) leads to
   \begin{equation}\label{46}
     1-\eta-|R'_{8}-R'_{15}+\eta|\sqrt{1-\frac{R'^2_{9}}{(R'_{8}+\eta)R'_{15}}}=0.
   \end{equation}
   Equation (\ref{46}) is a power equation of $\eta$. For any state on the curve CD described by Eq.(\ref{24e}),
    the coefficients of Eq.(\ref{46}) are determined by a single parameter $K$. The solution of (\ref{46}) is denoted as $\eta(K)$.
    The left side of Eq.(\ref{46}) reaches its local maximal value of 0 at $\eta(K)$. Hence, all the states on curve segment CD are triseparable. Criterion IV is necessary and sufficient when applied
    to curve segment CD (and similarly to HI in Fig.2 and DE in Fig.1).

    The value of $\eta(K)$ is equal to 0 at point C, with $K=0.6626275, v=0.7492394$, and $\alpha=8.900032$
    as determined for $\eta(K)=0$. If we further reduce $K$ (increase $\alpha$), we will obtain a negative $\eta(K)$.
   Hence, C is the end point for sufficiency of criterion IV.

\section{Conclusion}
   We proposed a road map for finding the separability criteria of multipartite entangled states.
The separability criteria are obtained with respect to a certain level of separability or entanglement depth.
For a hierarchy of state sets $\mathcal{S}_{1}\subset\mathcal{S}_{2}\subset\cdot\cdot\cdot\subset\mathcal{S}_{N}$,
there is a hierarchy of criterion sets $\mathcal{C}_{1}\subseteq\mathcal{C}_{2}\subseteq\cdot\cdot\cdot\subseteq\mathcal{C}_{N}$.
Each criterion set detects the separability of the corresponding state set necessarily and sufficiently.
The necessity and sufficiency of a criterion are related in the form of the eigenvalue and eigenvector of a certain matrix.
Thus, the route to the necessary condition indicates the route to the sufficient condition.

We developed the criterion set $\mathcal{C}_{2}$ for the triseparability of the highly symmetric four-qubit GHZ diagonal state set $\mathcal{S}_{2}$.
The criterion set $\mathcal{C}_{2}$ contains four criteria. All the criteria are necessary and sufficient in their application regions.
One of the criteria is just the criterion (the only criterion in set $\mathcal{C}_{1}$) that is necessary and sufficient for the triseparability of the generalized Werner state set $\mathcal{S}_{1}$.
The set $\mathcal{S}_{1}$ for generalized Werner states is a subset of the set $\mathcal{S}_{2}$
for highly symmetric GHZ diagonal states. Moreover, the EW set $\mathcal{W}_{2}$ for deriving $\mathcal{C}_{2}$ shares some properties with the EW set $\mathcal{W}_{1}$ for deriving $\mathcal{C}_{1}$.
We showed that $\mathcal{C}_{1}\subset\mathcal{C}_{2}$. Our criterion set $\mathcal{C}_{2}$ is also applicable to the four-qubit GHZ diagonal state set $\mathcal{S}_{3}$, although it is not sufficient for some of the states in $\mathcal{S}_{3}$.
Numerical calculation showed that $\mathcal{C}_{2}\subset\mathcal{C}_{3}$ (see Appendix E), where $\mathcal{C}_{3}$ is the necessary and sufficient criterion set for $\mathcal{S}_{3}$. Hence, we have $\mathcal{C}_{1}\subset\mathcal{C}_{2}\subset\mathcal{C}_{3}$.

With an explicit example, we showed that a larger criterion set for precisely detecting the entanglement of a larger state set would be developed more easily from a smaller criterion set for a smaller state set.
\section*{Acknowledgement}

Supported by the National Natural Science Foundation of China(Grant No£º11375152) and (partially) supported by National Basic Research Program of
China (Grant No. 2014CB921203) are gratefully acknowledged.

\section*{Appendix A: Proof of formula for $\widetilde{g}_{j}$}
The definition of $g_{1}$ yields $|g_{1}(\varphi_{1},\varphi_{2})|=[(K_{8}\cos\varphi_{1}\cos\varphi_{2}+K_{14}\sin\varphi_{1}\sin\varphi_{2})^2+(K_{10}\cos\varphi_{1}\sin\varphi_{2}\\
+K_{12}\sin\varphi_{1}cos\varphi_{2})^2]^{\frac{1}{2}}.$
Hence
\begin{eqnarray}\nonumber
  \widetilde{g}_{1}&=& \max_{\varphi_{1},\varphi_{2}}|g_{1}(\varphi_{1},\varphi_{2})|\\
   &=& \max_{\varphi_{1},\varphi_{2},\varphi_{3}}(K_{8}\cos\varphi_{1}\cos\varphi_{2}\cos\varphi_{3}\nonumber\\
   &&+K_{14}\sin\varphi_{1}\sin\varphi_{2}\cos\varphi_{3}\nonumber\\
   &&+K_{10}\cos\varphi_{1}\sin\varphi_{2}\sin\varphi_{3}\nonumber\\
   &&+K_{12}\sin\varphi_{1}cos\varphi_{2}\sin\varphi_{3}).\nonumber
\end{eqnarray}
According to Lemma 1 of \cite{Chen2017}, the maximization can be evaluated analytically. Hence the formula for $\widetilde{g}_{1}$ as a function of $K_{8},K_{10},K_{12},K_{14}$  follows.
We have similar result for $\widetilde{g}_{2}$.

\section*{Appendix B: Supplementary to criterion I }
   In (\ref{14}), we have set $x=\frac{\widetilde{g}_2}{\Lambda}=0$ to derive (\ref{16}) which leads to criterion I. If we set $x=\frac{\widetilde{g}_2}{\Lambda}>0$ in (\ref{11}),
we would get a smaller $\widetilde{\mathcal{L}}$. The minimal $\widetilde{\mathcal{L}}$ would be $\mathcal{L}_{min}=(|R_{7}|+\widetilde{R}_{1}+\widetilde{R}_{2})^{-1}$ when $x=1$. This is not true.
The reason is that the symmetric assumption (\ref{6y}) may be violated when we set $\widetilde{g}_2=\widetilde{g}_1$. When the symmetric assumptions are violated,
we have to calculate the largest eigenvalues of six different $\mathcal{M}$ matrices, then $\Lambda$ is the maximum of all the eigenvalues.

   Take an example of EW with $K_{8}=1,K_{10}=K_{12}=K_{14}=-1$, $K_{9}=x,K_{11}=K_{13}=K_{15}=-x $ with $x\geq0$ for partition $1|2|34$, we have $\widetilde{g}_1=1, \widetilde{g}_2=x$ for this partition.
Then $M_{8}=-M_{10}=-M_{12}=-M_{14}=\frac{1}{2}(1+x)$, $M_{9}=M_{11}=M_{13}=-M_{15}=\frac{1}{2}(x-1)$. If we interchange the second qubit and the third qubit in partition $1|2|34$,
we arrive at partition $1|3|24$. The parameters $M_{i}$ undergo interchanges $M_{9}\Leftrightarrow M_{10}, M_{13}\Leftrightarrow M_{14}$. For partition $1|3|24$, we have $\widetilde{g}_1=1+x, \widetilde{g}_2=0$.
Similarly for partitions $1|4|23$ and $2|4|13$, we have $\widetilde{g}_1=1, \widetilde{g}_2=x$; for partitions $2|3|14$ and $3|4|12$, we have $\widetilde{g}_1=1+x, \widetilde{g}_2=0$.
Notice that $\Lambda$ is the maximum of all these $\widetilde{g}_{1},\widetilde{g}_{2}$ in the six partitions. Hence $\Lambda=1+x.$ We may set $|M_{7}|=\Lambda$ and keep the assumption (\ref{6x}).
Then after taking optimization with respect to $x$, we have
\begin{equation}\label{47}\nonumber
  \mathcal{L}_{\min}=\frac{1}{|R_{7}|+\max{(\widetilde{R}_{1},\widetilde{R}_{2})}}.
\end{equation}
Here ${\widetilde{R}}_{j}=\max_{m=7,9,11,13}{|t_{m+j}|}$, with $(t_{9},t_{11},t_{13},t_{15})=(T_{9},T_{11},T_{13},T_{15})\Gamma$.
From (\ref{47}),the entanglement criterion for the triseparability follows:
\begin{eqnarray}\label{48}
\text{Criterion I':\qquad }\max_{i=1}^{16}|\rho_{i,17-i}|\leq\frac{1}{2}\min(\rho_{1,1}+\rho_{4,4}\nonumber\\
+\rho_{6,6}+\rho_{7,7},\text{\quad}\rho_{2,2}+\rho_{3,3}+\rho_{5,5}+\rho_{8,8}).\nonumber
\end{eqnarray}
It is a update version of criterion I. The right hand of the criterion can be substituted by
$\frac{1}{4}\min(\rho_{1,1}+\rho_{4,4}+\rho_{6,6}+\rho_{7,7}+\rho_{10,10}+\rho_{11,11}+\rho_{13,13}+\rho_{16,16}
,\text{\quad}\rho_{2,2}+\rho_{3,3}+\rho_{5,5}+\rho_{8,8}+\rho_{9,9}+\rho_{12,12}+\rho_{14,14}+\rho_{15,15})$.

\section*{Appendix C: Criterion for generalized Werner states}
   For generalized Werner states $\rho_{W}$,
we have EW with parameters $M_{i}$ described at the end of section \ref{section2}. Then $K_{8}=-K_{10}=-K_{12}=-K_{14}=2, K_{6}=-K_{7}=2$, all the other $K_{j}=0$.
It follows that $(\xi,\beta,\gamma,\delta)=\frac{1}{4}(K_{8},K_{10},K_{12},K_{14})\Gamma=(-1,0,0,0)$, hence $\xi\beta\gamma\delta=0$. Thus $\widetilde{g}_{1}$ is determined by the second line of the formula for $\widetilde{g}_{1}$.
We have $\widetilde{g}_{1}=2$. Together with $M_{7}=2$ and $\widetilde{g}_{2}=0$, we arrive at $\Lambda=\max(|M_{7}|,\widetilde{g}_{1},\widetilde{g}_{2})=2.$
The optimal EW is $\hat{W}=2\mathbf{I}-\hat{M}$. We have $Tr(\rho \hat{W})=2-\sum_{i=1}^{15}M_{i}R_{i}=2-(2R_{7}+R_{8}+R_{15}-6R_{9})\geq 0$. It leads to criterion I.
Applying it to generalized Werner states, we have the triseparable condition $p\geq \frac{1}{5} $. generalized Werner state with $p=\frac{1}{5}$ can be decomposed to tripartite separable states\cite{Chen2017}.
Thus criterion I is the necessary and sufficient criterion for generalized Werner states.

\section*{Appendix D: Details of the states in subsection \ref{AA}}
 In subsection \ref{AA},
  the triseparable state is a proper mixture of $\overline{\varrho}_{1}$ and $\rho'$ for the necessary and sufficient criterion I.
The explicit expression of $\overline{\varrho}_{1}$ is
    \begin{eqnarray}\label{27}
    \overline{\varrho}_{1}(\varphi_{1},\varphi_{2})=\frac{1}{16}\{IIII+\frac{1}{6}(IIZZ+IZIZ+IZZI+ZIIZ\nonumber\\
     +ZIZI+ZZII)+\frac{1}{2}(\cos^2\varphi_{+}+\cos\varphi_{+}\cos\varphi_{-})XXXX\nonumber\\
    -\frac{1}{6}(1+\sin^2\varphi_{+})(XXYY+XYXY+XYYX+YXXY\nonumber\\
     +YXYX+YYXX)+\frac{1}{2}(\cos^2\varphi_{+}-\cos\varphi_{+}\cos\varphi_{-})YYYY\}\nonumber
    \end{eqnarray}
 The fully separable state $\rho'$ is
  \begin{eqnarray}\nonumber
    \rho'=\frac{1}{16}[IIII+(1-\frac{\kappa}{6(1-\kappa)})(IIZZ+IZIZ\\
    +IZZI+ZIIZ+ZIZI+ZZII)+ZZZZ] \nonumber\\
    =[\frac{1}{2}-\frac{\kappa}{16(1-\kappa)}](|0000\rangle\langle0000|+|1111\rangle\langle1111|)\nonumber\\
    +\frac{\kappa}{48(1-\kappa)}(|0011\rangle\langle0011|+|1100\rangle\langle1100|+|0101\rangle\langle0101|\nonumber\\
    +|1010\rangle\langle1010|+|0110\rangle\langle0110|+|1001\rangle\langle1001|)\nonumber
  \end{eqnarray}
  The positive definiteness of $\rho'$ can be proven as follows. From (\ref{19b})(\ref{19c}) and (\ref{24a0}), we have $R_{8}+R_{15}=2(1-2p_{16})(1-\frac{4+6v}{\alpha})$. Using (\ref{30}) leads to $R_{8}+R_{15}=\frac{4(\alpha-7)}{\alpha}(1-2p_{16})\cos^2\varphi_{+}.$ From (\ref{28a})and  (\ref{28b}), we have $\kappa\cos^2\varphi_{+}=|R_{8}+R_{15}|$, hence $\kappa=\frac{4|\alpha-7|}{\alpha}(1-2p_{16})$. When $\alpha=9$ (straight lines AB in Fig.1 and Fig.2), we have $\kappa\leq \frac{8}{9}$. When $\alpha=5$ (straight line FG in Fig.2), the physical condition $p_{1}\geq 0$ leads to $p_{16}\geq \frac{2}{9}$, thus we also have $\kappa\leq \frac{8}{9}$. So that $\frac{1}{2}-\frac{\kappa}{16(1-\kappa)}\geq 0$, and $\rho'$ is positive definite.

The triseparable state for the necessity and sufficiency of criterion (\ref{21}) is a mixture of triseparable states $\overline{\varrho}_{23}$ and $\overline{\varrho}_{45}$, with
\begin{eqnarray}\label{34}\nonumber
      \overline{\varrho}_{23(45)}(\theta,\varphi)=\frac{1}{16}\{IIII\pm\frac{1}{6}(1+\cos^2\theta)(IIZZ+IZIZ \nonumber\\
      +IZZI+ZIIZ+ZIZI+ZZII)+\cos^2\theta ZZZZ \nonumber\\
      +\sin^2\theta[\cos^2\varphi XXXX+\sin^2\varphi YYYY\mp\frac{1}{6}(XXYY \nonumber\\
      +XYXY+XYYX+YXXY+YXYX+YYXX)]\}. \nonumber
    \end{eqnarray}

\section*{Appendix E: Criterion set $\mathcal{C}_{2}$ is not equal to $\mathcal{C}_{3}$}

We may wander if the criterion set $\mathcal{C}_{2}$ suffices for GHZ diagonal states, namely $\mathcal{C}_{2}=\mathcal{C}_{3}$,
where $\mathcal{C}_{3}$ is the necessary and sufficient triseparability criterion set for all four qubit GHZ diagonal states. The above example
indicates that $\mathcal{C}_{2}=\mathcal{C}_{3}$ is unlikely true. In some circumstances,
we should make use of the parameters $M_{i}$ which are not symmetric under qubit exchange. In fact, the following numerical calculation shows that
$\mathcal{C}_{2}\neq\mathcal{C}_{3}$. Since we know that $\mathcal{C}_{2}\subseteq\mathcal{C}_{3}$. Hence we have $\mathcal{C}_{2}\subset\mathcal{C}_{3}$.

The numerical example is $[R_{8},...,R_{15}]=[0.3255, -0.5260 $,\\$0.0739, 0.4046, -0.8764, -0.4321, -0.5037, 0.8752]$. Using symmetric $M_{i}$ (i=8,...,15)
(namely, equation (\ref{6y}) fulfills),we calculate $\Lambda_{sym}=\max(\widetilde{g}_{1},\widetilde{g}_{2})$,
 the minimum of the expression $\frac{\Lambda_{sym}}{\sum_{i=8}^{15}M_{i}R_{i}}$ with respect to symmetric $M_{i}$ is 0.6641. Using asymmetric $M_{i}$,
  we calculate $\Lambda_{asym}$ which is the maximum of $\widetilde{g}_{1}$ and $\widetilde{g}_{2}$ of all the six partitions, the minimum of the expression
 $\frac{\Lambda_{asym}}{\sum_{i=8}^{15}M_{i}R_{i}}$ with respect to asymmetric $M_{i}$ is 0.5347. Hence asymmetric EW (without condition (\ref{6y}))
  is better than symmetric EW (with condition (\ref{6y})) in detecting entanglement of some GHZ diagonal states.

\end{document}